\documentclass[10pt, twocolumn, article]{revtex4}

\usepackage{amsmath}
\usepackage{amssymb}
\usepackage{graphicx}

\input epsf
\input rotate

\def\bea{\begin{eqnarray}}
\def\eea{\end{eqnarray}}
\def\ben{\begin{equation}}
\def\een{\end{equation}}
\def\benu{\begin{enumerate}}
\def\enu{\end{enumerate}}

\def\n{n}
\def\lsim {\ifmmode {\buildrel<\over\sim}}

\def\sss{\scriptscriptstyle\rm}

\def\1var{(\bx_1...\bx\N)}

\def\half{\frac{1}{2}}

\def\br{{\bf r}}

\def\b1{{\bf 1}}
\def\bx{{x}}

\def\Hxc{_{\sss HXC}}

\def\N{_{\sss N}}

\def\sph_int{ {\int d^3 r}}

\def\infintd3r{ \int_{-\infty}^\infty d^3r\,}
\def\intd3r{ \int d^3r\,}

\def\laplace1d{\frac{d^2}{dx^2}}
\def\plaplace1d{\frac{d^2}{d{x'}^2}}

\def\padr2{\frac{\partial^2}{\partial r^2}}

\def\N{{\cal N}}
\def\a{{\alpha}}
\def\b{{\beta}}

\begin{document}

\author{Jonathan Nafziger}
\email{jnafzig@purdue.edu}
\affiliation{Department of Physics, Purdue University, 525 Northwestern Avenue, West Lafayette, IN 47907, USA}
\author{Qin Wu}
\email{qinwu@bnl.gov}
\affiliation{Center for Functional Nanomaterials, Brookhaven National Laboratory, Upton, NY 11973}
\author{Adam Wasserman}
\email{awasser@purdue.edu}
\affiliation{Department of Chemistry, Purdue University, 560 Oval Drive, West Lafayette, IN 47907, USA}
\affiliation{Department of Physics, Purdue University, 525 Northwestern Avenue, West Lafayette, IN 47907, USA}

\title{Molecular Binding Energies from Partition Density Functional Theory}






\date{\today}

\begin{abstract}
Approximate molecular calculations via standard Kohn-Sham Density Functional Theory are {\em exactly} reproduced by performing 
self-consistent calculations on isolated fragments via Partition Density Functional Theory [Phys. Rev. A {\bf 82}, 024501 (2010)]. 
We illustrate this with the binding curves of small diatomic molecules. We find that partition energies are in all cases qualitatively 
similar and numerically close to actual binding energies. We discuss qualitative features of the associated partition potentials.
\end{abstract}


\maketitle


\section{Introduction}

Kohn-Sham Density Functional Theory (KS-DFT) \cite{HK64,KS65} provides one of the most useful methods for calculating electronic 
properties of molecules and materials \cite{books}. The accuracy that can be achieved with modern approximations to the 
${\rm xc}$-functional \cite{PBE96,TPSS03,ZT08}, and the efficiency of numerical implemenations \cite{NWChem}, make of 
KS-DFT one of the workhorses of computational quantum chemistry, materials science, and nanotechnology.

A method has recently been proposed that {\em exactly} reproduces the results of approximate KS-DFT calculations via 
self-consistent calculations on isolated fragments \cite{EBCW10}. {\em Partition Density Functional Theory} (PDFT) is 
based on the density-Partition Theory (PT) of ref.\cite{CW07}, and is analogous to KS-DFT in that it establishes a map between 
the physical system of ground-state density $n(\br)$, thought of as a collection of interacting fragments, and an auxiliary 
system of $N_f$ {\em non-interacting} fragments with ensemble-ground-state-densities $\{n_\a(\br)\}$ subject to the density constraint
$\sum_\a^{N_f} n_\a(\br)=n(\br)$. The appeal of this formalism is two-fold: on the one hand, 
by optimally dividing a complex system into fragments it allows one to build a rigorous foundation for chemical 
reactivity theory \cite{CW03, CW06, CW07}. On the other hand, by solving the molecular Kohn-Sham equations in a different way, it 
focuses attention on quantities that are amenable to new, different approximations, potentially leading to linear-scaling algorithms
for large systems. In that spirit, PDFT is similar to embedded-DFT \cite{HC06,HPC11,GAMM10,GBM11}. One such quantity is the {\em partition potential}, 
$v_p(\br)$, a global molecular property (called reactivity potential in ref.\cite{CW07}, and equivalent in practice to the
crystal potential of ref.\cite{C91} and the embedding potential of ref.\cite{HPC11}).  
It is used as an additional external potential for each fragment 
so that the density constraint is satisfied.  If the density of the whole is known, as in the original formulation of PT, $v_p(\br)$ 
is simply the Langrage multiplier of the density constraint.   Without knowing the whole density, an iterative procedure for 
deriving $v_p(\br)$ was designed and illustrated with 1-D models \cite{EBCW10,ECWB09}.  
In this work we will show how this procedure is realized for real molecules. This step is of course essential to 
be able to  explore the promise of PDFT on both fronts mentioned before.

The partition potential is the functional derivative of the {\em partition energy}, $E_p[\{n_\a\}]$, with respect to any of the fragment densities
$n_\a(\br)$. This energy is defined as the difference between the molecular ground-state energy $E[n]$ and the sum of the 
fragment energies $\sum_\a^{N_f} E_\a[n_\a]$. 

 Kohn-Sham equations can be established so that approximations to $E_p[\{n_\a\}]$  yield definite predictions for the ground-state 
energy and density of the assembly. When the exact $E_p[\{n_\a\}]$ is employed implicitly by inversion of molecular KS equations \cite{EBCW10}, 
then the exact KS-DFT results are recovered without ever having to solve the direct problem for the assembly, but only for the fragments.

Analytical studies on one-dimensional 
models of heteronuclear diatomics \cite{ZW10} provide an early indication that the fragment dipoles obtained by PDFT are more 
adjusted to chemical intuition and more transferable than those obtained by other density-partitioning schemes, but more studies are
of course needed in real systems.

In this work, by employing the Wu-Yang algorithm \cite{WY03} for iterative inversion, we demonstrate convergence of the PDFT 
equations in small diatomic molecules, and discuss qualitative features of partition potentials and partition-energy binding curves 
for He$_2$, H$_2$, and LiH. We show that the partition energies and potentials are interesting quantities in themselves, as they
can be used as conceptual and interpretative tools.

First, we summarize the PDFT procedure in Sec.\ref{sec:methods}, providing details of our implementation. Convergence of the PDFT 
equations is demonstrated in Sec.\ref{sec:results} for the binding curves of He$_2$, H$_2$, and LiH, along with implications, qualitative 
features of patition potentials, and $E_p$-binding curves (in addition to actual binding curves). 
Concluding remarks are given in Sec.\ref{sec:conclusions}.

\section{Method}
\label{sec:methods}

For the simplicity of discussion, we consider a compound with only two parts ($A$ and $B$), but the method is equally applicable 
to any number of fragments.  We also limit ourselves to fragments with fixed integer number of electrons, as in related recent work 
on embedding-DFT \cite{GAMM10,GBM11,HPC11}, only briefly discussing 
the issue of chemical potential equalization and fractional electron numbers.

In PDFT, the total energy is expressed as
\begin{equation}
 \label{eq:pdft}
E[n]=E_A[n_A] + E_B[n_B] + E_p[n_A,n_B]
\end{equation}
where $n(\br)=n_A(\br)+n_B(\br)$, and a common functional for $E, E_A$ and $E_B$ is assumed.  The above equation can be viewed as a formal
 and exact definition of $E_p$.  
To minimize $E$ by variations of fragments' densities, which are built from their own sets of orbitals, we have the 
following Kohn-Sham equations:
\begin{equation}
\label{eq:ks}
\left[-\half\nabla^2+v_\a(\br)+v_p(\br)+v\Hxc[n_\a](\br)\right]\phi_i^\a(\br)=\varepsilon_i^\a\phi_i^\a(\br)
\end{equation}
Here, $\alpha$ is a fragment index, i.e $A$ or $B$ in this work.  
The partition potential $v_p (\br)$ is common to both fragments, 
thus has no $\alpha$-index.

$v_p(\br)$ could be derived explicitly if we knew the functional form of $E_p[\{n_\a\}]$.  Without an expression for $E_p$
as an explicit functional of the $\{n_\a\}$, it is also possible to derive $v_p (\br)$ 
through an iterative procedure, which was  first proposed in ref.\cite{EBCW10} and we reiterate here.

Suppose that we are at the beginning of the $k$-th iteration.  We obtain all fragment densities $n_{\alpha}^{(k)}$ by solving 
Eq.\ \ref{eq:ks}.  We then construct a total pro-molecule density as $\tilde{\n}^{(k)}(\br)=\sum\n_{\alpha}^{(k)}(\br)$.  
Because the effect of $v_p(\br)$ is to 
make $\tilde{n}(\br)$ the same as the true ground-state density of the whole system $n_s(\br)$, the difference between 
$\tilde{n}^{(k)}(\br)$ 
and $n_s(\br)$ should be used a guidance to update $v_p^{(k)}(\br)$.  For that, we do a constrained search to find the 
energy of $\tilde{n}^{(k)}$, i.e. 
\begin{equation}
E[\tilde{n}^{(k)}]=\min_{n \to \tilde{n}^{(k)}} E[n].
\end{equation}

We employ the direct optimization algorithm of Wu and Yang \cite{WY03}, as used in calculating the frozen density energy in a recently-developed 
density based energy decomposition analysis \cite{WAZ09}.   Thus we rewrite the above equation as
\begin{equation}
E[\tilde{n}^{(k)}]=E_v[\tilde{n}^{(k)}] + E\Hxc[\tilde{n}^{(k)}]+  \min_{\Psi \to \tilde{\n}^{(k)}} \left\{T_s[\Psi] + E_{\rm X}[\Psi] \right\}
\label{eq:erho}
\end{equation}
for a general hybrid functional, where $E_{\rm X}[\Psi]$ represents a fraction of the HF exchange energy calcualted from a Slater determinant 
$\Psi$ that is constrained to yield $\tilde{n}^{(k)}$.
At the end of this minimization, the effective potential for the molecular Kohn-Sham orbitals is
\begin{equation}
v_{\rm eff}(\br) = v_\a(\br)+v\Hxc[\tilde{n}^{(k)}](\br)-v_{\lambda}(\br)~~,
\label{eq:v}
\end{equation}
where $v_{\lambda}(\br)$ is just the Lagrange multiplier corresponding to the density constraint and is expanded by a linear 
combination of atom-centered Gaussian functions.  Because $v_{\lambda}(\br)$ is used to force the density of the whole system 
to be $\tilde{n}^{(k)}$,  its reverse should have the effect of making $\tilde{n}(\br)$ more like $n_s(\br)$.  That is: we can 
set $v_p(\br)=-v_{\lambda}(\br)$ and start the next iteration of fragment calculations.   In practice, we update $v_p(\br)$ as follows:
\begin{equation}
v_p^{(i)}(\br) =  v_p^{(i-1)}(\br) - \theta * v_{\lambda}^{(i)}(\br)~~,
\end{equation}
where $i$ is the iteration number, and $\theta$ is a damping factor between 0 and 1 used to control convergence.  In our 
calculation,
we have used $\theta=1$ or $\theta=0.25$.  The convergency criterion we use is $|E[\tilde{n}_k]-E[n_s]|<\epsilon$, 
where $\epsilon=10^{-6}$; this guarantees the converged energy is the same at the ground-state energy.  The alternative choice of 
$|E[\tilde{n}_k]-E[\tilde{n}_{k-1}]|<\epsilon$ gives essentially the same results.  

\section{Results}
\label{sec:results}

We demonstrate our calculations of $v_p(\br)$ with three simple examples of diatomic systems: He$_2$, H$_2$, and LiH.  In all 
calculations, Dunning's aug-cc-pvTz basis set is used for molecular orbitals.  

The counter-poise (CP) method is used to account for any Basis Set Superposition Error (BSSE).  
This approach is crucial in PDFT since $v_p(\br)$ adds features to the fragment's effective potential directly at the location of the other atom, precisely where the ghost basis functions are added \cite{Jensen}.

 The partition potential is expanded by atom-centered Gaussian
 functions, and each center has five $s$-type functions, with even-tempered exponents of $2^{n}, n=0, \pm 2, \pm 4$.   In 
the following discussion, we will use several energy terms.  Suppose $E_A^0$ and $E_B^0$ are the energies of the fragments 
with no influence of the partition potential; $E_A^p$ and $E_B^p$ are their energies with the converged partition potential; 
and $E_{AB}$ is the energy of the compound.  Therefore the binding energy is $E_{\rm bind}=E_{AB}-(E_A^0+E_B^0)$, and the 
partition energy is $E_p=E_{AB}-(E_A^p+E_B^p)$.  We also define the preparation energy as $E_{\rm prep}=(E_A^p+E_B^p)-(E_A^0+E_B^0)$, 
which is the energy increase associated with the deformation of fragments.  Clearly, $E_{\rm bind}=E_{\rm prep}+E_p$.

\subsection{Helium Dimer}

\begin{figure}[htbp]
\begin{center}

\includegraphics*[scale=0.65,clip=false,trim=0.9cm 0cm 0cm 0cm]{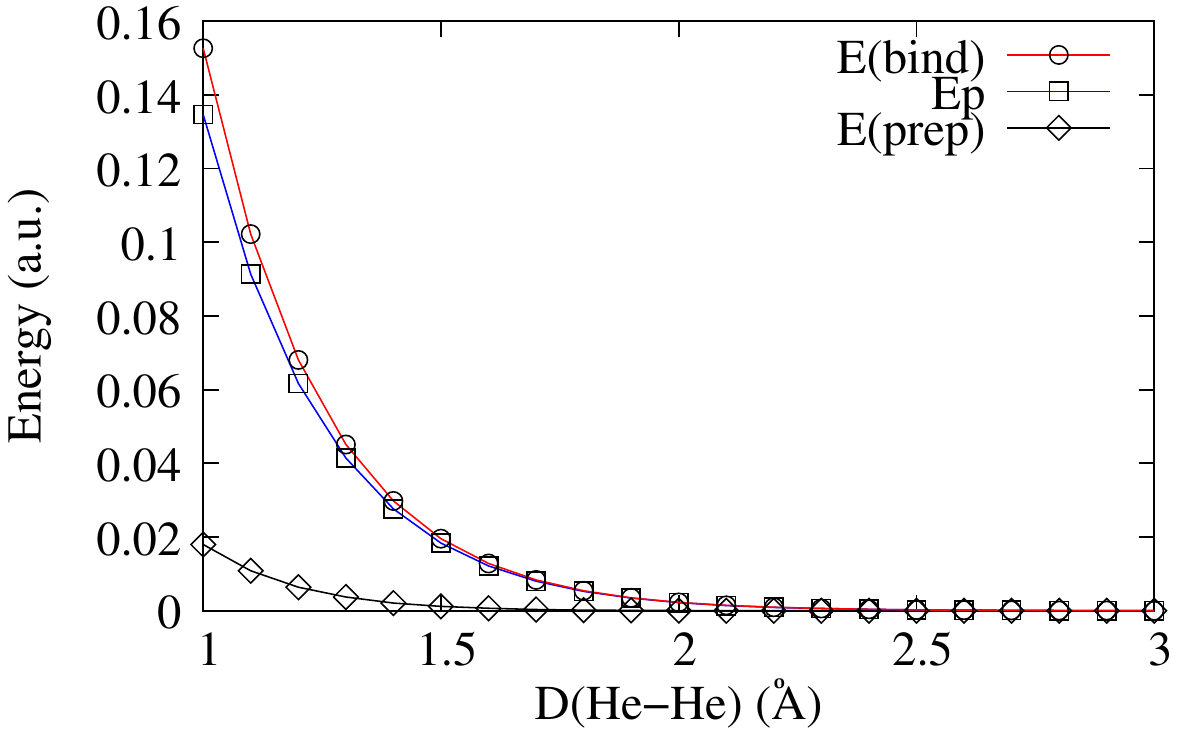}
\caption{The Hartree-Fock energies for He$_2$ at different internuclear distances.}
\label{fig:He2_Ep}
\end{center}
\end{figure}

\begin{figure}[htbp]
\begin{center}
\includegraphics*[scale=0.65,clip=false,trim=0.9cm 0cm 0cm 0cm]{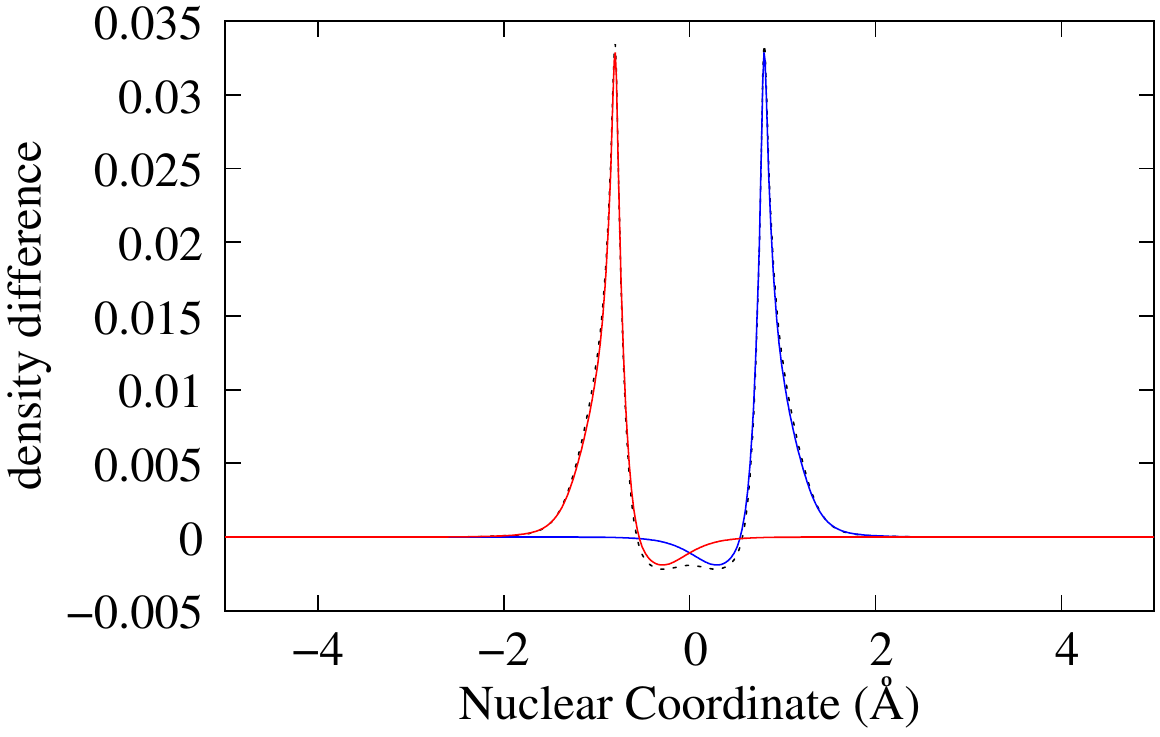}
\caption{The density differences in He$_2$ as compared to the original atoms along the line through both nuclei.  The total 
difference (dashed line) is the sum of the deformation in each atom (solid line).  The nuclei coordinates are $R=\pm$0.8 \AA. }
\label{fig:He2_drho}
\end{center}
\end{figure}

\begin{figure}[htbp]
\begin{center}
\includegraphics*[scale=0.65,clip=false,trim=0.9cm 0cm 0cm 0cm]{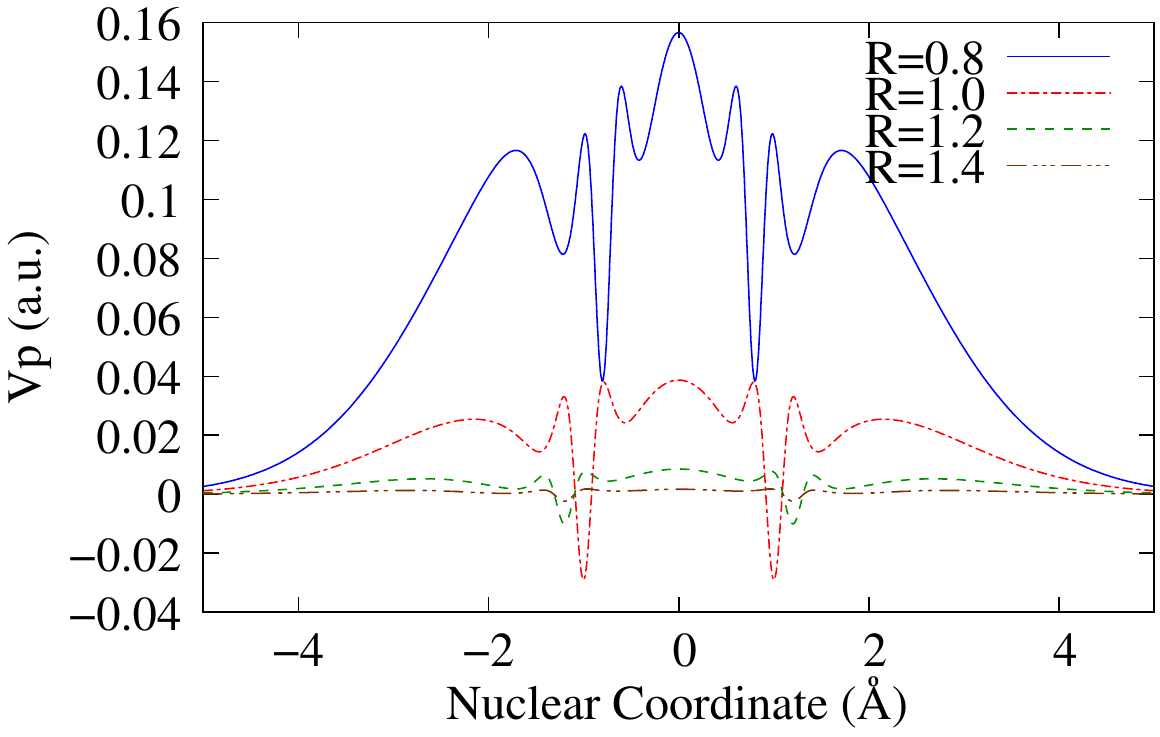}
\caption{The partition potentials for He$_2$ at different internuclear distances.  The nuclei are at $\pm R$.}
\label{fig:He2_Vp}
\end{center}
\end{figure}

Rare-gas dimers are known to be weakly bound due to van der Waals interactions, which are not accurately captured by most density-functional
approximations.  However, because our procedure is general and independent of the exchange-correlation functional, it is not critical 
to have the correct binding curve.  Instead, for a clear demonstration, we use Hartree-Fock exchange only, which is known 
to be purely repulsive between nonpolar closed-shell systems.   As shown in Fig.\ \ref{fig:He2_Ep}, the binding energy for He$_2$ 
is all positive and increases rapidly when the internuclear distance is shortened.   It also shows that the preparation energy 
is very small, which means the deformation in He atoms is small, as expected in this system, though it starts to grow when the 
atoms are too close to each other.  The repulsive nature of the interaction means that electron densities are pushed away from 
each other when the two He atoms are in close contact.   Thus  the internuclear region has a density decrease, as shown in 
Fig.\ \ref{fig:He2_drho}.  In PDFT, this density difference is achieved through deformation of each atom, due to the action of the 
partition potential.   In Fig.\ \ref{fig:He2_Vp} we plot $v_p$ along the internuclear axis at a few representative internuclear distances.  
Clearly, $v_p$ is most positive in the internuclear region, corresponding to the density deficiency.  The magnitude of $v_p$ 
decreases as the internuclear distance increases, until to a point that no $v_p$ is needed. 

It is notable that there are significantly more oscillations in the partition potential than in the density differences.  
Some of the oscillations are physical.  But there are at least two other possible reasons contributing to the oscillations in $v_p$.  
One is pathological with gaussian densities, as nicely explained by Schipper, Gritsenko and Baerends \cite{SGB97}. 
 The other is numerical and due to the fact that we expand $v_p$ in a finite basis set \cite{HBY07}.   We have used a small number of 
functions so as to limit the oscillations caused by the expansion.  However, we are unable to use non-gaussian densities yet,
 which makes it difficult to determine the nature of the oscillations.

\subsection{Hydrogen Molecule}

\begin{figure}[htbp]
\begin{center}

\includegraphics*[scale=0.65,clip=false,trim=0.9cm 0cm 0cm 0cm]{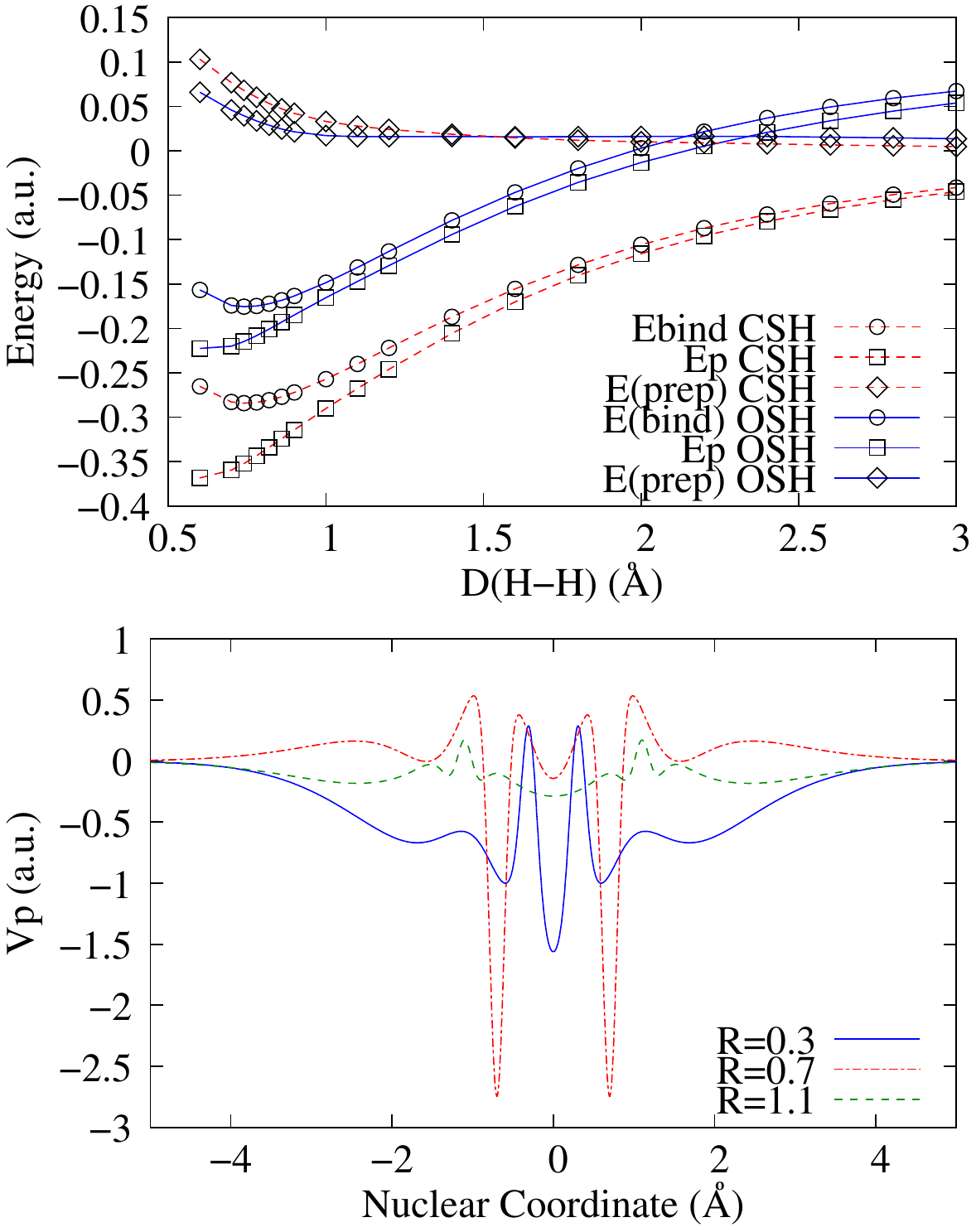}
\caption{Top: The B3LYP energies for H$_2$ at different interatomic distances.  The optimized bond length is $D=0.743$ \AA. Bottom: The partition potentials for H$_2$ at different internuclear distances.}
\label{fig:H2}
\end{center}
\end{figure}

For the covalently bonded molecule H$_2$, the natural choice of partition is to use two open-shell H (OSH) atoms.  
Because their spins are paired up in the molecule, we only consider the total charge density.  Mathematically one could also
 use half-occupied closed-shell H atoms (CSH) as the fragments, thus without polarizing the spin.  We study the energetics of 
both partitions as a function of the internuclear distance, using the B3LYP approximation to the exchange-correlation functional. 
 For the H$_2$ molecule, we only consider restricted Kohn-Sham (RKS) calculations.  It is well-known that a restricted calculation 
does poorly for large internuclear distances.  The erroneous behavior is evident from the binding energy curve when the OSH 
atoms are used as the reference.  As shown in Fig.\ \ref{fig:H2}, $E_{\rm bind}$ approaches a positive value instead of zero.  
On the other hand, when the CSH atoms are used as the reference, $E_{\rm bind}$ does go to zero.  However, it becomes too large 
at the optimal bond length.  The two binding curves are simply different by a constant shift, and this shift comes from the fact 
that OSH and CSH have different energies in the B3LYP approximation, while they should be degenerate with the exact functional 
\cite{CMY08}.

In PDFT, the differences in the choice of fragments will not matter if the partition energy can compensate for the difference 
and yield identical total energy.  In our case here, the two $E_p$ are indeed quite different.  However, the two $E_p$ curves differ 
more than by a simple constant shift.  The non-uniform difference can be appreciated by comparing the preparation energies.  
$E_{\rm prep}$ of OSH fragments is smaller at short internuclear distances than that of CSH fragments.  However, the latter 
goes to zero at long distances while the former does not.  At long distances, a restricted H$_2$ is essentially two half-occupied 
closed-shell H atoms, so the asymptotic behavior is not surprising.  But it is interesting to see that at short distances, the OSH 
fragments pay less penalty to make their densities resemble that of the molecule.

\subsection{Lithium Hydride}
As another example, we consider the heteronuclear LiH.  Within the formal partition theory, there is a unique choice of the
 fragments, with their chemical potentials equilibrated.  Achieving equilibration requires treating fragments with fractional 
number of electrons in the spirit of PPLB \cite{PPLB82}.  In that case, the number of electrons in a fragment is also a variable to be optimized.  
Because the partition potential will be different when the fragments change, the optimization of both the partition potential and 
the number of electrons is mutually dependent and has to be achieved simultaneously.  We will treat this complexity in the future.  
In this work, we simply use fixed fragments and derive the corresponding partition potential.  

\begin{figure}[htbp]
\begin{center}
\includegraphics*[scale=0.65,clip=false,trim=0.9cm 0cm 0cm 0cm]{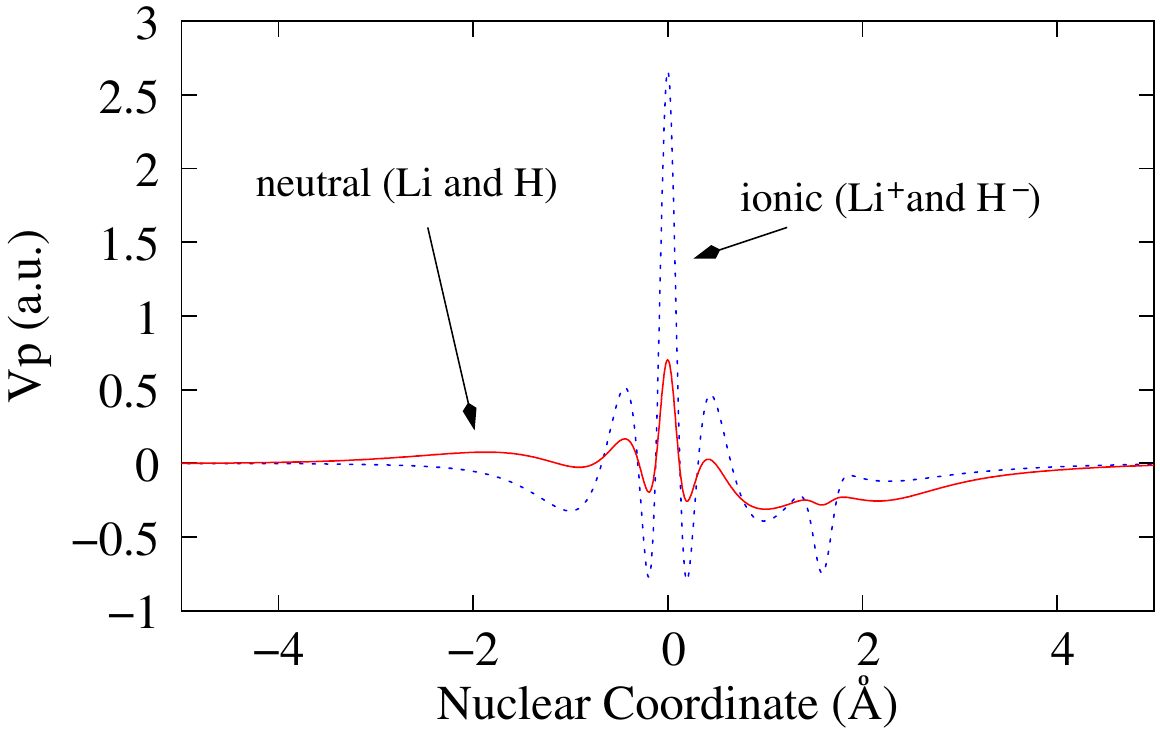}
\caption{The B3LYP partition potentials for LiH. The Li atom is at $x=0$ and H at $x=1.59073$ \AA.}
\label{fig:LiH}
\end{center}
\end{figure}

Without the optimal fragments, we consider all possible partitions.  For LiH, there are two possibilities.  First, we use 
neutral atoms.  Second, we use Li$^+$ and H$^-$.  We do the partition at the optimized internuclear distance of 1.59073 \AA.  
 For the neutral partition, $E_{\rm prep}=0.053257$ a.u. and $E_p=-0.146407$ a.u.  For the ionic partition, $E_{\rm prep}=0.034395$ a.u. and $E_p=-0.300161$ a.u.  The larger partition energy in the ionic case could be the result of Coulomb attraction.  
However, the preparation energy is smaller for the ionic partition, suggesting the LiH bond is closer to an ionic bond than 
a covalent bond.  What is surprising is that the partition potential for the ionic case looks much stronger than the neutral 
one (Fig.\ \ref{fig:LiH}), despite causing less distortion in fragment's energies.

\section{Concluding Remarks}
\label{sec:conclusions}

Without having to solve directly the KS equations for the total external potential, we have shown how the PDFT algorithm of ref.\cite{EBCW10} provides in practice the same answers via {\em fragment}-KS equations. In addition, this method yields fragment densities, fragment energies, and a partition potential that is shared by all fragments such that the sum of their densities reproduces the correct total density.

Although no physical meaning can be attached to a parition potential beyond the one implied by its definition (i.e. that it is the potential 
common to all fragments such that the sum of the fragment densities equals the total molecular density), 
some generic features of partition potentials seem to go in line with chemical intuition:  they are positive when the 
interaction between fragments is repulsive (case of He$_2$ within Hartree-Fock), and their average magnitude is larger 
when the interaction between fragments is stronger. Similarly, the strength of the interaction between fragments is 
loosely measured by the magnitude of the partition energy. No such conclusion can be drawn for the preparation energy, 
however, as shown for the case of LiH where a somewhat larger preparation energy is associated with a much smaller 
partition potential (neutral vs. ionic partition).  But the preparation energy can tell us about the {\em character} 
of the bond, an aspect that we plan to study further in future work. The case of LiH also highlights the need to go 
beyond integer numbers of electrons in our implementation of PDFT.

PDFT calculations also allow us to look at the dissociation problem from a different angle. For example, we found that open-shell 
fragments in H$_2$ are preferred at short inter-nuclear separations in the sense that they pay less penalty to make their 
densities resemble that of the molecule, but close-shell fragments are preferred at long separations. The respective 
preparation energies cross near the Coulson-Fischer point.

Finally, we point out that from weak (He$_2$) to relatively strong (H$_2$) chemical bonds, partition energies are qualitatively similar 
to actual binding energies, and numerically close to them (i.e. preparation energies are {\em small} in the cases studied). This similarity of $E_p$-curves 
to their corresponding binding curves suggests that  approximations of $E_p[\{n_\a\}]$ as explicit functionals of 
the $\{n_\a\}$ might be very useful for practical computations. Not only would they provide a direct way to obtain the partition 
potentials by functional differentiation, circumventing the need of expensive inversion steps; sensible approximations would 
also lead to energies that are close to actual binding energies.  This is analogous to what happens in KS-DFT, whose success is 
largely due to the fact that the sum of KS orbital energies is typically close to actual ground-state energies in chemical applications. 

\section{Acknowledgments}
We acknowledge support from the Office of Basic Energy Sciences, U.S. Department of Energy, 
under grant No.DE-FG02-10ER16196. This research was carried out in part at the Center for Functional Nanomaterials, Brookhaven National Laboratory, which is supported by the U.S. Department of Energy, Office of Basic Energy Sciences, under Contract No. DE-AC02-98CH10886.


\begin{thebibliography}{0}

\bibitem{HK64}
P. Hohenberg and W. Kohn, Phys. Rev. {\bf 136}, B 864 (1964).

\bibitem{KS65}
W. Kohn and L.J. Sham, Phys. Rev.  {\bf 140}, A 1133 (1965).

\bibitem{books}
R.M.  Dreizler  and  E.K.U.  Gross,    {\em Density   Functional   Theory }
    (Springer-Verlag, Berlin, 1990);
R.G. Parr and W. Yang, {\em Density-Functional  Theory  of  Atoms  and Molecules}
(Oxford, New York, 1989);
R.M. Martin, {\em Electronic Structure} (Cambridge University Press, Cambridge, 2004). 


\bibitem{PBE96}
J.P.~Perdew, K.~Burke, and M.~Ernzerhof, Phys. Rev. Lett. {\bf 77}, 3865 
(1996); {\bf 78}, 1396 (1997) (E).

\bibitem{TPSS03}
J. Tao, J.P. Perdew, V.N. Staroverov and G.E. Scuseria, Phys. Rev. Lett. {\bf 91}, 14640 (2003).

\bibitem{ZT08}
Y. Zhao and D.G. Truhlar, Theor. Chem. Acc. {\bf 120}, 215 (2008).

\bibitem{NWChem}
M. Valiev, E.J. Bylaska, N. Govind, K. Kowalski, T.P. Straatsma, H.J.J. van Dam, D. Wang, J. Nieplocha, E. Apra, T.L. Windus, W.A. de Jong, 
Comput. Phys. Commun. 181, 1477 (2010)

\bibitem{EBCW10}
P. Elliott, K. Burke, M.H. Cohen, and A. Wasserman,
Phys. Rev. A {\bf 82}, 024501, (2010).

\bibitem{CW07}
M.H. Cohen and A. Wasserman, J. Phys. Chem. A {\bf 111}, 2229
(2007).

\bibitem{CW03}
M.H. Cohen and A. Wasserman, Isr. J. Chemistry {\bf 43}, 219 (2003).

\bibitem{CW06}
M.H. Cohen and A. Wasserman, J. Stat. Phys. {\bf 125}, 1125 (2006).

\bibitem{HC06}
C. Huang and E.A. Carter,
J. Chem. Phys. {\bf 125}, 084102 (2006).

\bibitem{HPC11}
C. Huang, M. Pavone, and E.A. Carter, 
J. Chem. Phys. {\bf 134}, 154110 (2011).

\bibitem{GAMM10}
J.D. Goodpaster, N. Ananth, F.R. Manby, and T.F. Miller III, J. Chem. Phys. {\bf 133}, 084103 (2010).

\bibitem{GBM11}
J.D. Goodpaster, T.A. Barnes, and T.F. Miller III, arXiv:1102.4028v1 (2011).

\bibitem{C91}
P. Cortona, Phys. Rev. B {\bf 44}, 8454 (1991).

\bibitem{ECWB09}
P. Elliott, M.H. Cohen, A. Wasserman, and K. Burke, J. Comp. Theor. Chem. {\bf 5}, 827 (2009).

\bibitem{ZW10}
Y. Zhang and A. Wasserman, J. Chem. Theory Comput. {\bf 6}, 3312 (2010).

\bibitem{WY03}
Q. Wu and W. Yang, J. Chem. Phys. {\bf 118}, 2498 (2003).


\bibitem{WW93}
T.A. Wesolowski and A. Warshel,
J. Phys. Chem. {\bf 97}, 8050 (1993).

\bibitem{GSL94}
S.R. Gadre, R.N. Shirsat, and A.C. Limaye, J. Phys. Chem. {\bf 98}, 9165 (1994).

\bibitem{KIAN99}
K. Kitaura, E. Ikeo, T. Asada, T. Nakano, and M. Uebayasi, Chem. Phys. Lett. {\bf 313}, 701 (1999).

\bibitem{EM02}
T.E. Exner and P.G. Mezey, J. Phys. Chem. A {\bf 106}, 11791 (2002).

\bibitem{ZXZ03}
D.W. Zhang, Y. Xiang, and J.Z.H. Zhang, J. Phys. Chem. B {\bf 107}, 12039 (2003).

\bibitem{LLF05}
S.H. Li, W. Li, and T. Fang, J. Am. Chem. Soc. {\bf 127}, 7215 (2005).

\bibitem{HMK05}
L.L. Huang, L. Massa, and J. Karle, Int. J. Quantum Chem. {\bf 103}, 808 (2005).

\bibitem{DC05}
V. Deev and M.A. Collins, J. Chem. Phys. {\bf 122}, 154102 (2005).

\bibitem{BL06}
R.P.A. Bettens and A.M. Lee, J. Phys. Chem. A {\bf 110}, 8777 (2006).

\bibitem{RS10}
J. \u{R}ez\'{a}\u{c} and D.R. Salahub, J. Chem. Theory Comput. {\bf 6}, 91 (2010).

\bibitem{G99}
S. Goedeker, Rev. Mod. Phys. {\bf 71}, 1085 (1999).

\bibitem{WAZ09}
Q. Wu, P.W. Ayers, and Y. Zhang,
J. Chem. Phys. {\bf 131}, 164112 (2009).

\bibitem{Jensen}
F. Jensen, {\em Introduction to Computational Chemistry}, Wiley (1998).

\bibitem{SGB97}
P.R.T. Schipper, O.V. Gritsenko, and E.J. Baerends, Theor. Chem. Acc. {\bf 98}, 16 (1997).

\bibitem{HBY07}
T. Heaton-Burgess, F. A. Bulat, and W. Yang, Phys. Rev. Lett. {\bf 98}, 256401 (2007).

\bibitem{CMY08}
A.J. Cohen, P.Mori-S\'{a}nchez, and W. Yang,
J. Chem. Phys. {\bf 129}, 121104 (2008).

\bibitem{PPLB82}
J.P. Perdew, R.G. Parr, M. Levy, and J.L. Balduz, Jr.,  Phys.  Rev. Lett.  {\bf 49}, 1691 (1982).

\end{thebibliography}

\end{document}
%